\documentclass[aps,twocolumn,reprint,floatfix,letterpaper,nofootinbib]{revtex4-1}

\pdfoutput=1

\newcommand\cfig[1]{\vtop{\vskip-0.15in\hbox{\includegraphics[height=0.25in]{#1}}}}
\renewcommand\textcite[1]{\Ref{#1}}

\begin{document}

\begin{titlepage}

\title{Choreographed entangle dances: topological states of quantum matter}

\author{Xiao-Gang Wen}
\affiliation{Department of Physics, Massachusetts Institute of
Technology, Cambridge, Massachusetts 02139, USA}

\begin{abstract} 
For a long time, we thought that only symmetry breaking can give rise to
different phases of matter.  If there was no symmetry breaking, there would be
no pattern and it would be featureless.  But now we realize that, for quantum
matter at zero temperature, even symmetric disordered liquids can have
features, which give rise to topological phases of quantum matter.  Some of the
topological phases are highly entangled (\ie have topological order), while
others are weakly entangled (\ie have SPT order). In this article, we will
briefly introduce those new zero-temperature states of matter and their amazing
emergent properties. We will also discuss their significance in unifying some
most basic concepts in nature.

\end{abstract}

\pacs{}

\maketitle

\end{titlepage}

{\small \setcounter{tocdepth}{1} \tableofcontents }

\section{Condensed matter goes topological}

\begin{figure}[b]
\centerline{
\includegraphics[scale=0.4]{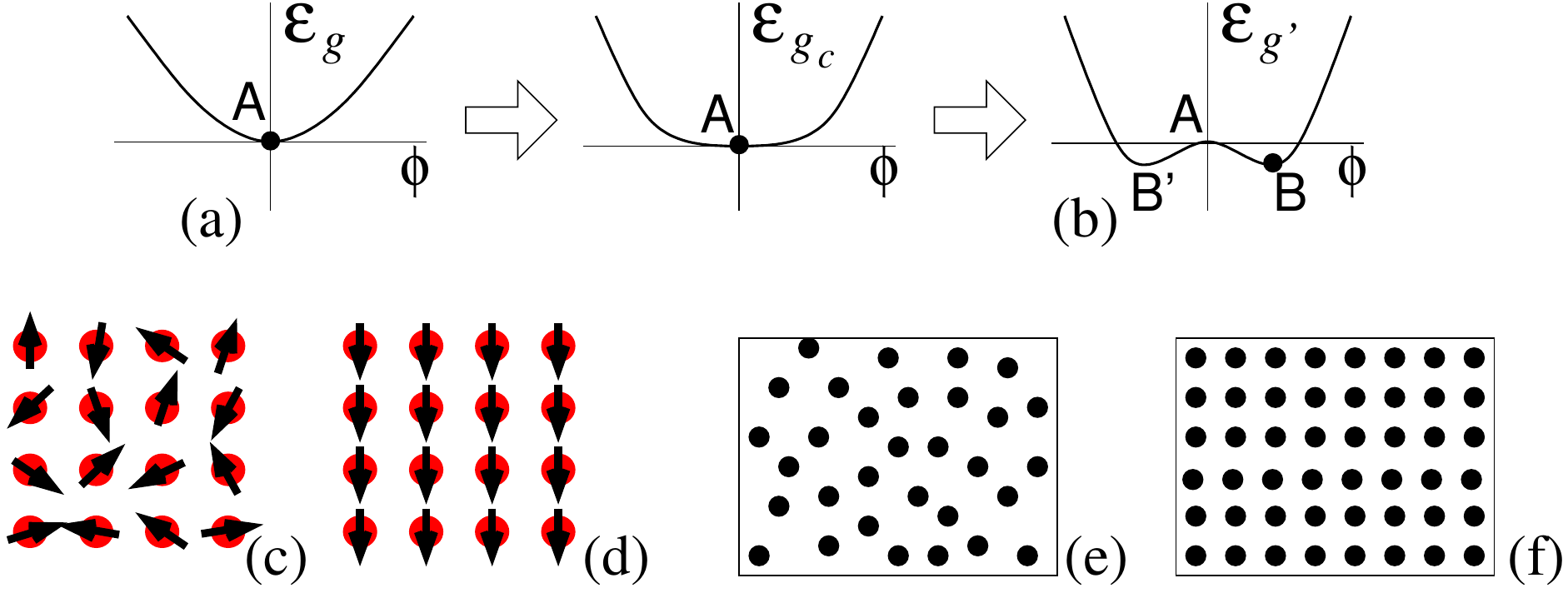}
}
\caption{ (a,c,e) Disordered liquid states that do not break any symmetry.
(b,d,f) Ordered states that spontaneously break some symmetries.  For example,
the energy function $\veps_g(\phi)$ has a symmetry $\phi \to -\phi$:
$\veps_g(\phi)=\veps_g(-\phi)$.  However, as we change the parameter $g$, the
minimal energy state (the ground state) some times respects the symmetry
(a,c,e), and other times have to settle into one that does not respect the
symmetry (b,d,f).  (d) is a ferromagnetic spin order and (f) is a crystal
order.
} \label{symmbrk}
\end{figure}

Since the 1980's, the study of topological phases of quantum matter slowly
became more and more active.  It now becomes a main stream in condensed
matter physics.  But what are topological phases?  Why people are
interested in topological phases?  We know that quantum matter refers
to states of matter at zero temperature, which can have many different kinds of
phases.  Topological phases are one class of those zero-temperature phases,
that appear featureless and have non-zero energy gaps.  The energy gap implies
that, like inert gases, topological phases hardly respond to any external
perturbations.  Featureless and inert, that sounds really boring. 

But in fact, topological phases of quantum matter represent a new unexplored
world in condensed matter physics.  Many amazing new phenomena, such as
emergent gauge interaction, emergent Fermi and non-Abelian statistics, are
discovered.  Some of those new phenomena are beyond wildest imaginations and
were thought to be impossible.  Those new topological phenomena have a deep
impact on different fields of physics and mathematics, such as condensed matter
physics, quantum information science, high energy particle physics, algebraic
topology and tensor category theory. In my opinion, the topological phases of
quantum matter and the associated many-body entanglement represent a second
quantum revolution in physics.

\section{The era of broken symmetry}

Our world is very rich. One aspect of its richness is reflected in the
existence of many different phases of matter.  It turns out that those rich
phases of matter can be understood from a symmetry point of view. For example,
a liquid has randomly distributed atoms. It has all the symmetries, since it
remains the same after we displace and/or rotate it arbitrarily.  Having all
the symmetry, liquids are featureless.  In contrast, a crystal does not have
all the symmetry.  It remains unchanged only when we displace it by a
particular set of distances (integer times of lattice constant). So a crystal
has only discrete translation symmetry.  The phase transition between a liquid
and a crystal is a transition that reduces the continuous translation symmetry
to the discrete symmetry. Such a change in symmetry is called ``spontaneous
symmetry breaking'' (see Fig.  \ref{symmbrk}).  The rich beautiful
crystal structures actually come from the partial breaking of the translation
and rotation symmetries.

Landau theory generalizes the above picture to describe any phases and any
phase transitions \cite{LL58}.  It points out that different phases are
different only because they have different symmetries in the organizations of
the constituent particles (such organizations are called orders).  As a
material changes from one phase to another, what happens is that the symmetry
of the organization of the particles changes.  Having such a comprehensive
theory that describes all phases of matter, one started have a feeling that the
condensed matter theory had come to its mature end.

\section{New world beyond symmetry breaking}

\subsection{Disordered liquids are not featureless}

According to the Landau symmetry breaking theory, the rich beautiful patterns
in phases of matter actually come from symmetry breaking.  If there was no
symmetry breaking, such as in disordered liquids, then there would be no
pattern and the state would be featureless.  But, in late 1980s, it became
clear that even disordered liquids can have features.  In a study of high $T_c$
superconductors \cite{KL8795,WWZcsp} of a 2-dimensional disordered spin liquid
-- ``chiral spin liquid'' -- was discovered.  Chiral spin liquid is
characterized by its absence of any spin order (see Fig. \ref{symmbrk}c) and
its perfect heat conducting edge (since all edge excitations move in the same
direction).  It was quickly realized that there can be many different chiral
spin liquids with exactly the same symmetry \cite{Wrig} but different numbers
of heat conducting edge modes \cite{Wedge}. So symmetry alone is not enough to
characterize and distinguish different chiral spin liquids.  This means that
the chiral spin liquids must contain a new kind of order that is beyond the
usual symmetry description.  The proposed new kind of order was named
``topological\footnote{The term ``topological'' was motivated by the low energy
effective theory of the chiral spin liquids which is a Chern-Simons (CS) gauge
theory \cite{WWZcsp} -- a topological quantum field theory (TQFT)\cite{W8951}.}
order''\cite{Wrig} 

Unfortunately, the chiral spin liquid has not been realized in experiments\footnote{The chiral spin liquid is realized numerically in Heisenberg model on
Kagome lattice with $J_1$-$J_2$-$J_3$ coupling \cite{HC14072740,GS14121571}.}.
But, people have discovered many different fractional quantum Hall (FQH) phases
at semiconductor interface under strong magnetic field.  FQH liquids have a
property that an electric field will induce a current density in the transverse
direction: $j_y=\si_{xy} E_x$.  It is an amazing discovery that the Hall
conductance $\si_{xy}$ of a FQH liquid is precisely quantized as a rational
number $\nu=\frac{p}{q}$ if we measure it in unit of $\frac{e^2}{h}$:
$\si_{xy}=\nu\frac{e^2}{h} $ \cite{KDP8094,TSG8259}.  Different quantized
$\si_{xy}$ correspond to different FQH phases.  Just like the chiral spin
liquids, those different FQH phases all have the same symmetry and cannot be
distinguished by symmetry-breaking.  This suggests that FQH liquids may also
contain the new topological orders.

\subsection{What is the essence of FQH liquids?}

C. N. Yang once pointed out that the BCS theory of fermionic superfluid and
superconductor
capture the essence of the  superfluid and superconductor, but
what is this essence?  To address this question, he developed the theory of
off-diagonal long range order \cite{Y6294} which reveal the essence of
superfluid and superconductor. In fact long range correlation of local order
parameter is the essence of any symmetry breaking order.  

Similarly, Laughlin's theory \cite{L8395} captures the essence of the FQH
effect, but what is this essence?  It turns out that the essence is not long
range order. Actually, the essence hidden in chiral spin liquids and FQH effect
is so new that it does not even have a name.  So we are free to call it
``topological order''.  In last 20 some years, the theory of topological order
was developed trying to understand what is this hidden essence.  Our
exploration is a journey into an unknown world, which is surprisingly rich and
fascinating.

\subsection{What is topological order?}

In the above discussion, we only refer topological order by what it is not: it
is not symmetry breaking order.  But this is not enough, we must define it by
what it is. In physics, to propose a new concept, we must define it via
physical properties that can be measured in experiments and/or numerical
calculations.  Furthermore, those measurable properties must be robust against
any small perturbations that can break any symmetries, in order to define new
orders beyond symmetry.  As a comparison, we note that superfluid order is
defined/characterized by zero viscosity and vortex quantization, that are
robust against any small perturbations that do not break the
particle-number-conservation symmetry.  Thus, superfluid order is an order
protected by particle-number-conservation symmetry.  

\begin{figure}[tb]
\hfil
\includegraphics[height=0.9in]{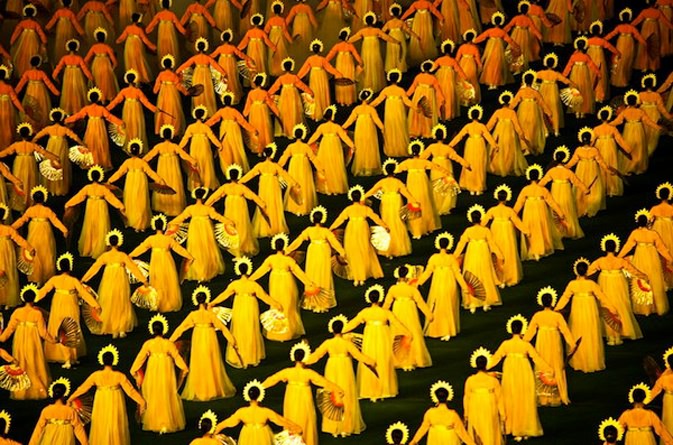}
~~~~
\hfil
\hfil
\includegraphics[height=0.9in]{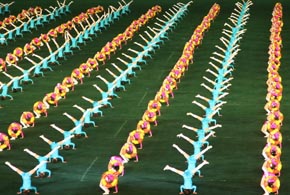}~~~\\
\hfil
(a) ferromagnetic order
\hfil
~~~~~~~~~~~~ (b) stripe order ~~~~~\\[2mm]
\hfil
\includegraphics[height=1.0in]{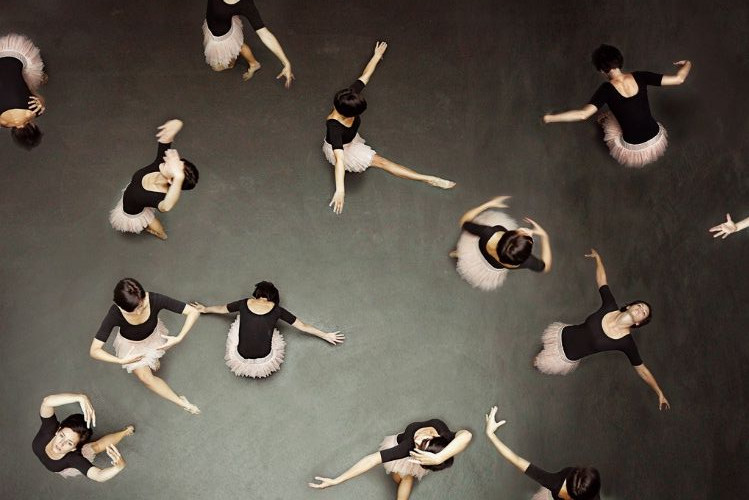}
\hfil
\hfil
\includegraphics[height=1.0in]{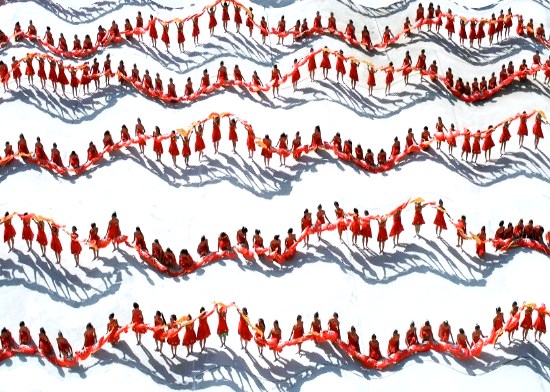}\\
\hfil
~~~~~~~~~~ (c) FQH liquid 
\hfil
\hfil
~~
~~~~~ 
(d) String liquid (spin liquid)
\caption{
(a,b) The static patterns for the symmetry breaking orders
(\ie product states).
(c,d) The dancing patterns for the topological orders:
global correlated group dances.
}
\label{dance}
\end{figure}

How to define/characterize topological order?  Influenced by Landau symmetry
breaking theory, people still try to use order parameter and long range
correlation to characterize the essence of FQH liquids
\cite{GM8752,ZHK8982,R8986},  which result in the Ginzburg-Landau-Chern-Simons
effective theory of quantum Hall states.  But the order parameter and long
range correlation used in the characterization are not physical.  In fact, all
the local physical operators have a short-ranged correlations in FQH liquids.
So to reveal the  essence of FQH liquids, we must think in a new direction.

Motivated by research in chiral spin state \cite{Wrig}, we identified new
topological properties that can reveal the essence of FQH liquids:
(1) the \emph{ground state degeneracy}\cite{Wrig,WNtop} on torus (and other spaces with non-trivial topology);
(2) the \emph{non-Abelian geometric phase}\footnote{For an explanation of
non-Abelian geometric phase, see \Ref{WZ8411}.}
 of those degenerate ground states
\cite{Wrig} (which form a representation of modular group $SL(2,\Z)$);
(3) the \emph{gapless edge modes} \cite{H8285,Wcll}.
 We showed that all those properties are robust against any small perturbations
(even those that break all the symmetries) \cite{WNtop,Wcll}. Thus those robust
topological properties play the role of order parameter which allow us to
define the notion of topological order macroscopically.  

The notion of topological order can also be defined microscopically.  First, we
reminder readers that a \emph{product state} is a many-body state formed by a
fixed pattern of local quantum state on each site, such as the
anti-ferromagnetic state: $|\down\up\down\up\down\up\down\>$. Product states
are \emph{unentangled} states.  A spin flip
$|\down\up\down{\color{red}\down}\down\up\down\>$ represents an excitation above the
anti-ferromagnetic ground state.  If such an excitation costs a finite energy,
then we say the ground state have a finite energy gap.  The notion of product
state is opposite to that of topological order in the sense that a gapped
product state has no topological order.  More generally: a topologically
ordered state is a \emph{gapped ground state}\footnote{More precisely, a
topologically ordered state is a \emph{gapped liquid state}, a notion
introduced in \Ref{ZW1490,SM1403}.} that cannot be continuously deformed into a
product state smoothly without closing the energy gap (\ie without phase
transition).  Two topologically ordered states belong to the same phase (\ie
have the same topological order) if they can deform into each other smoothly
without phase transition. From the above definition, one can show that all the
states with no topological order actually belong to the same phase, since the
deformation mentioned above can deform any product state into any other product
state.  

We note that the deformations that do not close energy gap can only modify the
entanglement over a short distance. Thus we can also call the states with no
topological order as short-range entangled states, and states with topological
order as long-range entangled states \cite{CGW1038}.  We see that topological
order is nothing but pattern of long range entanglement.  Long range
entanglement is the essence of topological order, as well as the essence of FQH
liquids and chiral spin liquids.

\section{Topological orders as choreographed quantum dances}

Topological order is a very new concept that actually describes quantum
entanglement in many-body systems.
Such a concept is very remote from our daily experiences and it is hard to have
an intuition about it.  Here we would like describe topological order through
some intuitive pictures.

\subsection{Static pattern versus dancing pattern}
\label{dance}

The product states (which correspond to symmetry breaking states) are
characterized by their fixed static patterns, such as
$|\down\up\down\up\down\up\down\>$ (see Fig. \ref{dance}a,b).  In contrast,
topologically ordered ground states are superpositions of many different
configurations, such as different spin configurations
$|\down\up\down\up\down\up\down\>+|\down\up\up\up\down\down\down\>+\cdots$.
Such kind of states are also referred as having strong quantum fluctuations.
In the following, we will use dancing to gain an intuitive picture of
topological order (\ie pattern of quantum fluctuation or quantum entanglement).


A topological order is described by a global dance (see Fig. \ref{dance}c,d),
where every particle (or spin) is dancing with every other particle (or spin)
in a very organized way: (a) all spins/particles dance following a set of
\emph{local} dancing ``rules'' trying to lower the energy of a \emph{local}
Hamiltonian.  (b) the local dancing
``rules'' enforce  a global dancing pattern, which corresponds to
the topological order
(\ie long-range entanglement).

For example in FQH liquid, 
the electrons dance according to the following local
dancing rules:\\ 
(a) electron always dances anti-clockwise which implies that
the electron wave function only depend on the electron coordinates $(x,y)$
via $z=x+\ii y$.\\
(b) each electron always takes exactly ``three steps'' to dance around any other
electron, which implies that the phase of the wave function changes by $6\pi$
as we move an electron around any other electron.\\
The above two local dancing rules enforce a global dancing pattern which
corresponds to the Laughlin wave function \cite{L8395} $\Phi_\text{FQH} = \prod
(z_i-z_j)^3$.  Such a collective dancing gives rise to the topological order
(or long range entanglement) in the filling fraction $\nu=1/3$ FQH liquid (\ie
with Hall conductance $\si_{xy}=\nu\frac{e^2}{h}$).

\begin{figure}[tb]
\centerline{
\includegraphics[height=.9in]{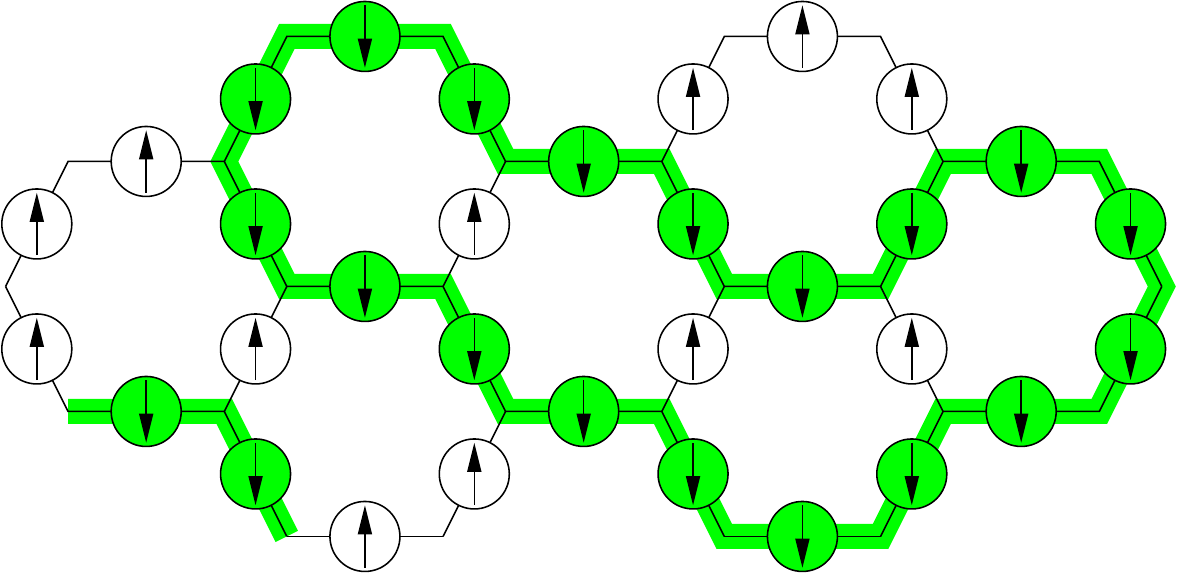}
}
\caption{
The strings in a spin-1/2 model.
In the background of up-spins, the down-spins form closed strings.
}
\label{strspin}
\end{figure}

In additional to FQH liquids, some spin liquids\footnote{Spin liquids refer to
ground states of quantum spin systems that do not \emph{spontaneously} break
any symmetry.} also contain topological orders \cite{RS9173,Wsrvb}, where
the spins dance according to  the follow local dancing rules:\\
(a) Down spins form closed strings with no open ends (see Fig.
\ref{strspin}).\\ 
(b) Strings can otherwise move and reconnect freely. \\ 
The
global dance formed by the spins following the above dancing rules gives us a
quantum spin liquid which is a superposition of all closed string
configurations:\cite{K032} $|\Phi_\text{string}\>=\sum_\text{all string
pattern} \left | \cfig{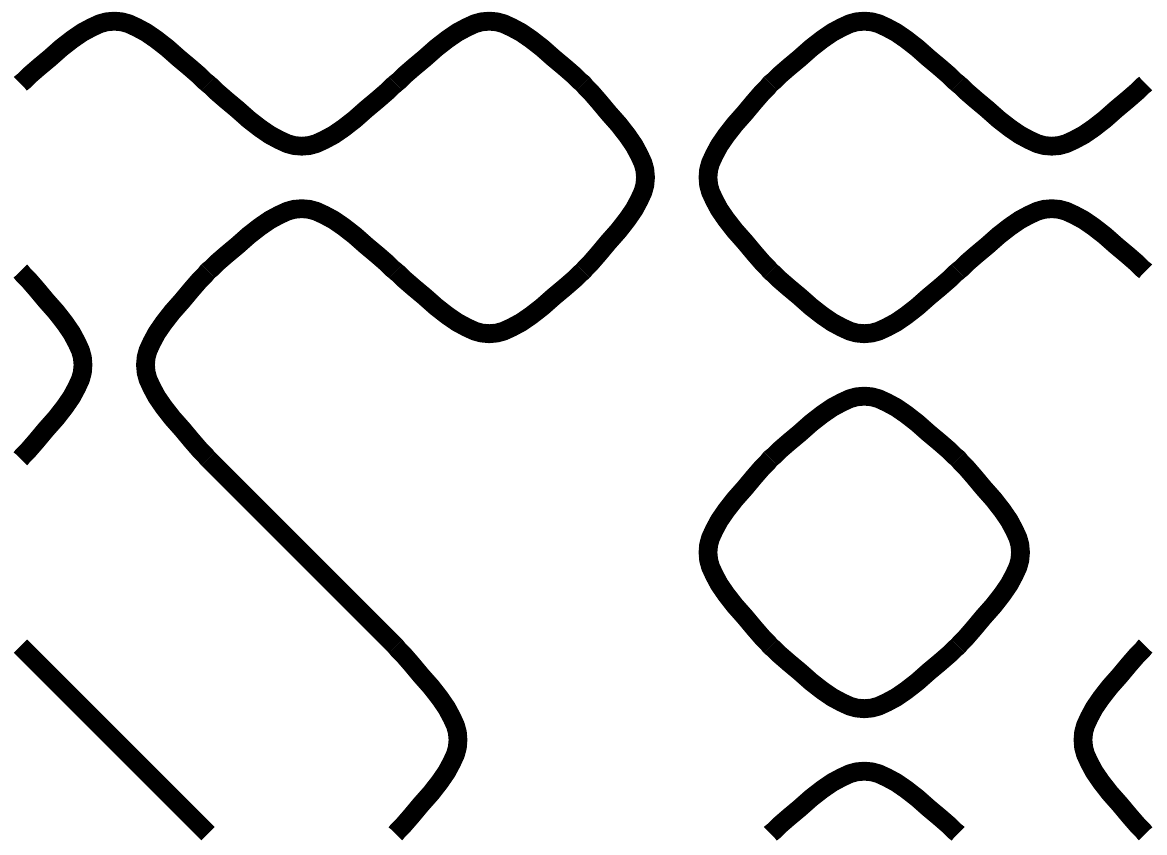}\right \> $.  Such a state is a string
condensed state \cite{LWstrnet}, which gives rise to the simplest topological
order --  $Z_2$ topological order \cite{RS9173,Wsrvb}.  

\subsection{Topological orders in flat land}

After having an intuitive picture of topological order in terms of quantum
dances, we like to ask, which topological orders have been observed in
experiments?  Which topological orders are possible?  What properties do they
have?

\subsubsection{Fractionalization without fragmentation}

In addition to the filling fraction $\nu=1/3$ FQH liquid, many other FQH
liquids with $\nu=2/5,2/3,3/7,\cdots$ were also discovered in experiments.
Those FQH states are described by more complicated dancing pattern.  For
example, the electrons in $\nu=2/5$ FQH liquid have the following dancing
pattern: (1) First, the electrons spontaneously separate into two components;
(2) all the electrons dances anti-clockwise; (3) each electron always takes
exactly three steps to dance around any other electron in the same component,
but takes exactly two steps to dance around any other electron in the other
component.  We can use a symmetric integer matrix to encode such a dancing
rule: $ K=\big({ 3\ 2 \atop 2\ 3 }\big)$. Similarly, the $\nu=2/3$ FQH liquid
is described by $ K=\big({ 1\ 2 \atop 2\ 1 }\big)$, while the $\nu=3/7$ FQH
liquid is described by a 3-by-3 $K$-matrix since its electrons separate into
three components.

Knowing the dance pattern encoded by $K$-matrix, we can calculate all the
topological properties of the FQH state from $K$.
For example the quantized Hall conductance is given by $\si_{xy}=\v q^T K^{-1}
\v q \frac{e^2}{h}$, where $\v q^T=(1,1,\cdots,1)$ describes the charge of the
electron in each component.  As if the precisely quantized Hall conductance is
not amazing enough, the FQH liquids (and other topologically ordered states)
also have particle-like excitations above the ground state, that may carry
fractional charges \cite{L8395} and fraction statistics \cite{ASW8422,H8483}.
We know that FQH liquids are formed by electrons, which all carry integer
charge.  It is hard to believe the appearance of fractionally charged
excitations without the electrons getting fragmented into smaller pieces.  But
this unbelievable prediction has been confirmed by noise measurement in
experiments \cite{dRH9762,SGJ9726}.

The fractional charge is hard to imagine, but the fraction statistics is even
more exotic.  We know that exchanging two identical particles will change their
quantum amplitude by a phase factor $\ee^{\ii\th}$.  If the particles are
bosons, then $\th=0$.  If the particles are fermions,  then $\th=\pi$.  In
nature, bosons and fermions are the only two types of elementary particles, and
their compositions are always bosons or fermions.  However, the excitations in
FQH liquids can give us totally new types of particles where $\th$ is not zero
nor $\pi$ \cite{LM7701,W8257}.  Such new type of particles is said to have
Abelian fractional statistics, or simply fractional statistics.  All those
fractionalization phenomena are not due to the fragmentation of electrons, but
due to the long range entanglement between electrons in topologically ordered
states.

As a result, the fractionalization is determined by the $K$-matrix that
describes the dancing pattern (\ie the pattern of long range entanglement)
\cite{WNtop,WZ9290}.  First, let us view each column of the $K$-matrix as a
vector, and those vectors span a lattice. The unit cell of such lattice contain
det$(K)$ number of integer vectors $\v l$.  Those integer vectors $\v l$ happen
to label det$(K)$ distinct types of fractionalized excitations.  The fractional
charge of the excitation is then given by $Q=\v q^T K^{-1} \v l$, and the
fractional statistical angle  by $\th =\pi \v l^T K^{-1} \v l$. 

It turns out that our simple dancing-step picture of topological order and long
range entanglement is very powerful and comprehensive:  all 2+1D Abelian
topological orders (\ie with only Abelian fractional statistics) can be
described by such kind of dance. In other words, all 2+1D Abelian topological
orders are classified by symmetric integer $K$-matrices \cite{WZ9290}.  This
dancing-step picture can be generalized even further which leads to a
pattern-of-zeros theory for FQH liquids \cite{SL0604,BKW0608,WW0808}, that can
even describe non-Abelian FQH liquids with non-Abelian statistics.

\subsubsection{Even degrees of freedom can be fractionalized}

But what is non-Abelian statistics?  We know that the Abelian statistics is
determined by the phase as we exchange two identical particles.  However, if
two particles have internal degrees of freedom, which leads to a degeneracy
$D$, then exchanging the two particles while leads to a $D\times D$ unitary
matrix.  In this case, we say the particles have a non-Abelian statistics
\cite{W8413,GMS8503,W8951,K062}.

We see that the key to have non-Abelian statistics is the degeneracy from the
internal degrees of freedom of the particles.  If the total degeneracy is $D_N$
for $N$ identical particles, then internal degrees of freedom of each particle
is given by $d=D_N^{1/N}$ in $N\to \infty$ limit.  For example, for spin-1/2
electron, $D_N=2^N$ and $d=2$. Thus the internal degrees of freedom of an
electron is 2, which corresponds to a 2-dimensional vector space in quantum
theory. We will call $d$ the quantum dimension of the particle. But for
particle with non-Abelian statistics, $d$  may not even be an
integer,\footnote{Even though $N$ and $D_N$ are integers, $d={\text{\footnotesize lim} \atop
N\to \infty} D_N^{1/N}$ may not be integer.} as if the internal degrees of
freedom of the particle are described by a vector space with a non-integer
dimension $d$. In this case, we say that the particle has a fractional degrees
of freedom! 

Non-Abelian statistics, in particular its fractional degrees of freedom, is so
strange, it hard to imagine such a thing can appear in condensed matter systems
formed by simple electrons and atoms.  But the quantum entanglement between
electrons is a very powerful ``creator'', it can even make such an
impossibility to happen.

A concrete realization of non-Abelian statistics in FQH liquids was first
proposed in \textcite{Wnab,MR9162}. One of the proposed FQH liquid at $\nu=2/3$
is given by the wave function \cite{Wnab}
$\Psi_{\nu=\frac{2}{3}}(\{z_i\})=[\chi_2(\{z_i\})]^3$, where $\chi_n(\{z_i\})$
is the many-electron wave function with $n$ filled Landau levels.  It has a
type $SU(3)_2$ 
non-Abelian statistics (\ie Fibonacci anyon), with quantum dimension
$d=\frac{1+\sqrt 5}{2}$ (\ie log$_2 d = 0.694$ qubits). Such $SU(3)_2$
non-Abelian statistics also appear in $Z_3$ parafermion FQH state
\cite{RR9984}.

The degeneracy from non-integer quantum dimension $d$ cannot be viewed as local
degrees of freedom associated with each particle, despite we call it ``the
internal degrees of freedom'' of the particle. Such a name, in this aspect, is
a little misleading.  In fact, those degrees of freedom are distributed between
well separated non-Abelian particles and are intrinsically non-local. As a
result, the quantum information carried by those degenerate states is immune
from the decoherence by environment which interact locally with each particle.
The degeneracy from non-Abelian particles is the same type of topological
degeneracy as the ground state degeneracy of topological order on torus. They
both are robust against any local perturbations, including any perturbations
acting on the particles.  So we can use non-Abelian topological order to
perform topological quantum computation \cite{F0028,FLZ0205,K032,K062}. In
particular, $SU(3)_2$ non-Abelian topological order can perform universal
topological quantum computation \cite{FLZ0205}.

A simpler non-Abelian FQH liquid at $\nu=1/2$ is given by
$\Psi_{\nu=\frac{1}{2}}(\{z_i\})=\chi_1(\{z_i\})[\chi_2(\{z_i\})]^2$
\cite{Wnab} or by $\Psi_{\nu=\frac{1}{2}} = \cA(\frac{1}{z_1-z_2}
\frac{1}{z_2-z_3} \cdots)\prod (z_i-z_j)^2$ \cite{MR9162}.  They both have type
$SU(2)_2$ non-Abelian statistics, with quantum dimension $d=\sqrt 2$ (a half
qubit!). Such non-Abelian particle is incorrectly called Majorana fermion by
some.  The experimentally realized $\nu=5/2$ FQH state is likely to be such
$SU(2)_2$ non-Abelian FQH state \cite{WES8776,DM08020930,RMM0899}.  But
unfortunately, $SU(2)_2$ non-Abelian topological order cannot perform universal
topological quantum computation \cite{FLZ0205}.

\subsubsection{String liquid in spin liquid: emergence of gauge theory}

\begin{figure}[tb]
\centerline{
\includegraphics[height=2.2in]{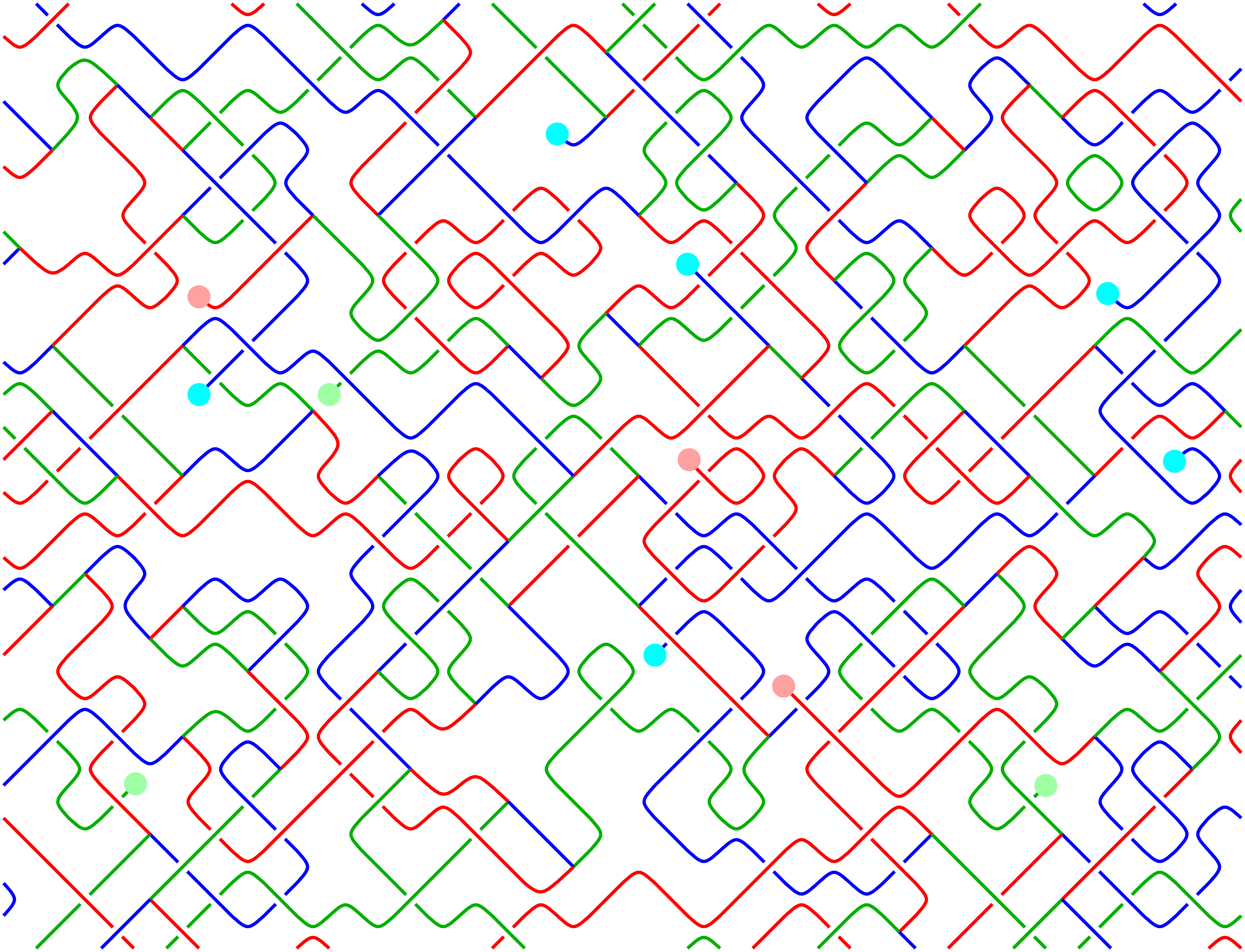}
}
\caption{
The color represents different types of string, which can join in a certain way
to form a string-net.  A sting-net liquid is a superposition of the above
string-nets. It can give rise to emergent (non-Abelian) gauge theory,
emergent non-Abelian statistics (in 2+1D) and emergent fermions (in 3D). It
unifies gauge interaction and Fermi statistics.  It provides an unified origin
for light and electrons, as well as other elementary particles. 
}
\label{stringnetW}
\end{figure}

FQH liquids and chiral spin liquids discussed above can not have time-reversal
symmetry.  However, there are topological orders with time-reversal symmetry,
such as the $Z_2$ topological order defined by its emergent $Z_2$-gauge theory
at low energy.  The $Z_2$ topological order (and the emergent $Z_2$-gauge
theory) was first proposed to exist in 2+1D spin liquid with frustrations
\cite{RS9173,Wsrvb}.  Later, a numerical calculation confirmed its existence in
a closely related quantum dimmer model on triangular lattice \cite{MS0181}.  If
we break the spin rotation symmetry, the $Z_2$ topological order can be
realized by an exactly soluble model -- the toric code model \cite{K032}.  The
toric code model reveals that the $Z_2$ topological order happen to be the
topological order in the string dance of non-oriented loops described in
section \ref{dance}.  Unlike FQH states whose boundary is always gapless, those
time-reversal symmetric topological orders 
can have boundaries that are gapped \cite{KK11045047}.  

In addition to the above theoretical realizations, the $Z_2$-topological order
\cite{RS9173,Wsrvb}
may be realized by Herbertsmithite (spin-1/2 on Kagome lattice) \cite{HMS0704},
as suggested by recent experiments \cite{FL151102174,HL151206807}.  The early
numerical calculation of \textcite{YHW1173} suggested the spin-1/2 Heisenberg
model on Kagome lattice to be gapped, but the details of the results are
inconsistent with $Z_2$-topological order, which led people to suspect that the
model is gapless. A more recent numerical calculation suggests the model to
have a $Z_2$-topological order with long correlation length (10 unit
cell length) \cite{MW160609639}, while several other calculations suggest
gapless $U(1)$ spin liquid ground states
\cite{JR161002024,LX161004727,HP161106238}.

We like to point out that the string dancing picture for the $Z_2$ topological
order and emergent $Z_2$ gauge theory can be generalized by allowing strings to
have more types and by allowing three strings to join in a certain way (see
Fig. \ref{stringnetW}).  This gives rise to  string-net condensed state
\cite{LWstrnet}, which can leads to  non-Abelian topological orders and
emergent non-Abelian gauge theory. In fact, the strings behave like the
``electric'' flux of the gauge theory and the string density wave corresponds
to non-Abelian gauge field that gives rise to gauge bosons.  The ends of
strings correspond gauge charges, which may carry non-Abelian statistics in
2+1D or Fermi statistics in 3+1D..

Such a string-net construction is also very powerful and comprehensive: it can
give rise to \emph{all} 2+1D topological orders with gappable boundaries
\cite{FNS0428,LWstrnet}. Such a construction has a mathematical root in unitary
fusion category theory \cite{ENO0562} and is closely related to the Turaev-Viro
invariant of 3-dimensional manifolds \cite{TV9265}.  To be more precise,
unitary fusion categories classify 2+1D topological orders with gappable
boundaries.

\section{Even disordered product states are not featureless}

For a long time, we thought that the symmetry breaking in the product states
describes all possible phases of matter.  Now we realize the existence of
long-range entangled states, which leads to rich topologically ordered phases.
If we consider only product states and assume no symmetry breaking, we would
expect the corresponding symmetric product states to be trivial.  However, this
naive guess turns out to be wrong.  If the Hamiltonian has a symmetry, even
when its ground state is a product state that does not spontaneously break any
symmetry, such ground state can still be non-trivial.  This is the most
``trivial'' non-trivial state of quantum matter.

\subsection{Haldane phase for integer spin chain}

Due to quantum fluctuations, the ground state of antiferromagnetic spin-1/2
chain does not break the $SO(3)$ spin-rotation symmetry.
However, it \emph{almost} breaks the symmetry in the sense that spin-spin
correlation has a slow algebraic decay (similar non-decaying long-range order)
and the chain is gapless (as if having a Goldstone mode).
For spin-$S$ chain with $S>1/2$, the quantum fluctuations are weaker than the
spin-1/2 chain.  So people believe that a spin-$S$ chain also \emph{almost}
breaks the $SO(3)$ symmetry, which is also  gapless and have an algebraic
decaying spin-spin correlation.

But in 1983, Haldane pointed \cite{H8364} out that the spin chain is actually
gapped and the spin-spin correlation has an exponential decay when the spin is
integer.  
The gapped ground state of integer spin chain is called Haldane phase.  At that
time, people did not distinguish even-integer-spin chain from odd-integer-spin
chain, and believed the Haldane phase for both even- and odd-integer spin chain
to be a trivial disordered phase, just like the product state formed by spin-0
on each site.

\subsection{Odd-integer and even-integer are different: the essence of
odd-integer Haldane phases}

In fact, only even-integer spin chains behave like the product states formed by
spin-0 on each site.  The odd-integer spin chains actually give rise to a
non-trivial state!

\begin{figure}[tb]
\begin{center}
\includegraphics[scale=0.45]{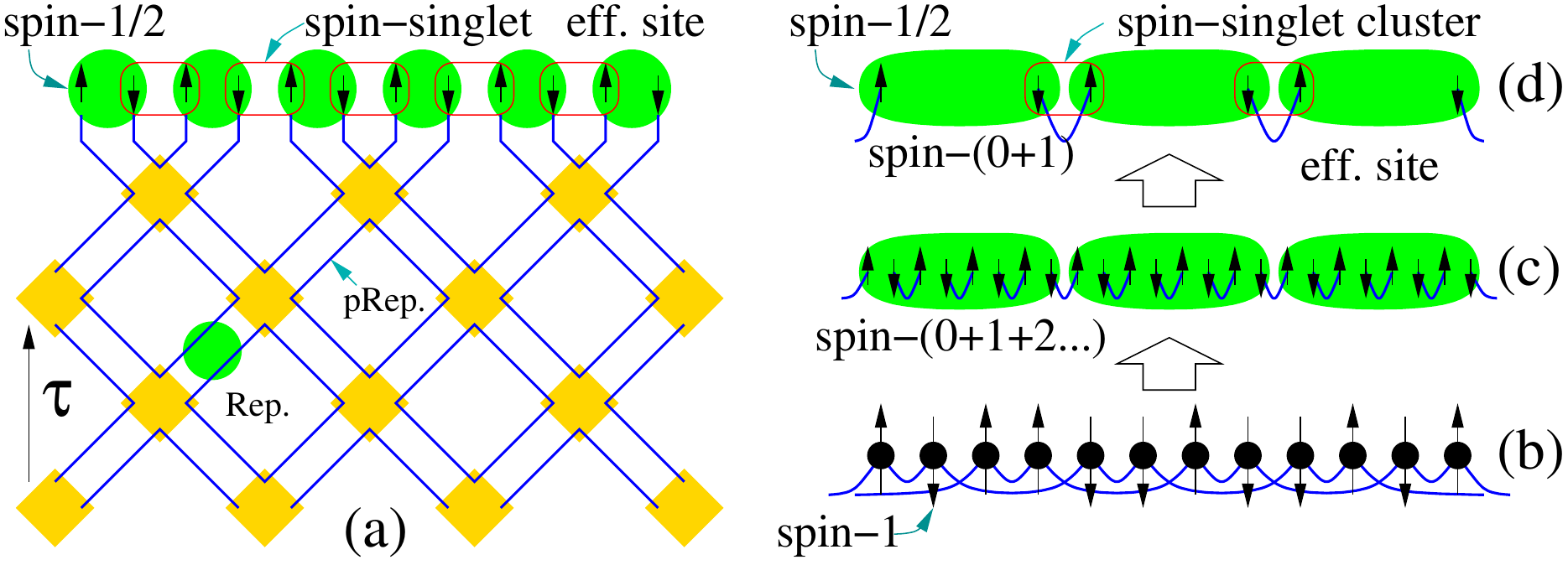}
\end{center}
\caption{
(a) Corner-double-line tensor obtained after coarse-graining the space-time
tensor network and its corresponding coarse-grained wave function.  (b,c,d) The
coarse-graining process in terms of ground state wave function.  The
entanglement structure is described by the blue lines.  The coarse-grained wave
functions have the same entanglement structure in (a) and (d).
} 
\label{cdlstate} 
\end{figure}

What is the essence of Haldane phase for odd-integer spins?  In \Ref{GW0931},
the spin-1 chain was studied using a tensor network renormalization approach.
It was found that the tensor network, describing the space-time spin
fluctuations, flows to a so called corner-double-line tensor under the coarse
graining of the network (see Fig. \ref{cdlstate}a).  To understand the physical
meaning of corner-double-line structure, let us describe the coarse-graining
process from the point of view of ground state wave function.

The ground state of spin-1 chain (see Fig. \ref{cdlstate}b) is formed by
spin-1's on each site.  We may group a number of sites into an effective site.
When the effective site is large enough, the direct entanglement between two
spin-1's (represented by the blue curves) can only appear between the
neighboring effective sites.  Then we can use an unitary transformation acting
within an effective site to simplify the entanglement within each effective
site (see Fig. \ref{cdlstate}c).  After removing the degrees of freedom that
are entangled only within each effective site, we obtain a simplified
coarse-grained wave function (see Fig.  \ref{cdlstate}d) which corresponds to
the corner-double-line structure.  We note that, in the coarse-grained wave
function, each effective site has four states: spin-0 $\oplus$ spin-1, which
can be viewed as two spin-1/2's: spin-1/2 $\otimes$ spin-1/2.  We also see that
the coarse-grained wave function is a product state of spin-singlets (see Fig.
\ref{cdlstate}a,d).  This seems confirm that the Haldane phase is a trivial
product state formed by spin-0's.

However, the coarse-grained wave function is not the  product state formed by
spin-singlets \emph{on each site}, but the  product state formed by
spin-singlets \emph{between sites} (see Fig. \ref{cdlstate}).  Furthermore, the
spin-singlets between sites are formed by spin-1/2's \cite{GW0931} which are
not representation of $SO(3)$, but projective representations of $SO(3)$
\cite{PBT1039,CGW1107}.  This feature and the associated corner-double-line
structure is robust against any perturbations that preserve of $SO(3)$
symmetry.  In other words, the product state formed by spin-singlets on each
site and the product state formed by spin-singlets between sites belong to two
different phases, provided that the spin-singlets are formed by projective
representations of $SO(3)$ (\ie half-integer spins). In this case, we cannot
deform one into the other without encounter gap closing phase transition, if
the deformation preserve the $SO(3)$ symmetry.  Thus the  Haldane
phase of spin-1 chain is actually a new phase of matter robust against any
symmetry preserving perturbations despite it is a product state that does not
break any symmetry.  This new phase of matter was named SPT phase (which stands
for symmetry protected trivial phase if one stresses its nature of short-range
entanglement, or  symmetry protected topological phase if one stresses its
nature of beyond symmetry breaking).

We see that an essence of spin-1 Haldane phase is the corner-double-line
structure in coarse-grained tensor network (see Fig. \ref{cdlstate}a), or
entangled clusters between sites without inter-cluster entanglement in
coarse-grained wave function (see Fig.  \ref{cdlstate}d).  \Ref{PBT1039} also
pointed out that Haldane phase has degenerate entanglement spectrum which
describes the intra-cluster entanglement.  The fact that the intra-cluster
entanglement is between projective representations of the symmetry is the other
essence of spin-1 Haldane phase (see Fig. \ref{cdlstate}a,d). 

\begin{figure*}[tb]
\centerline{
\includegraphics[height=1.9in]{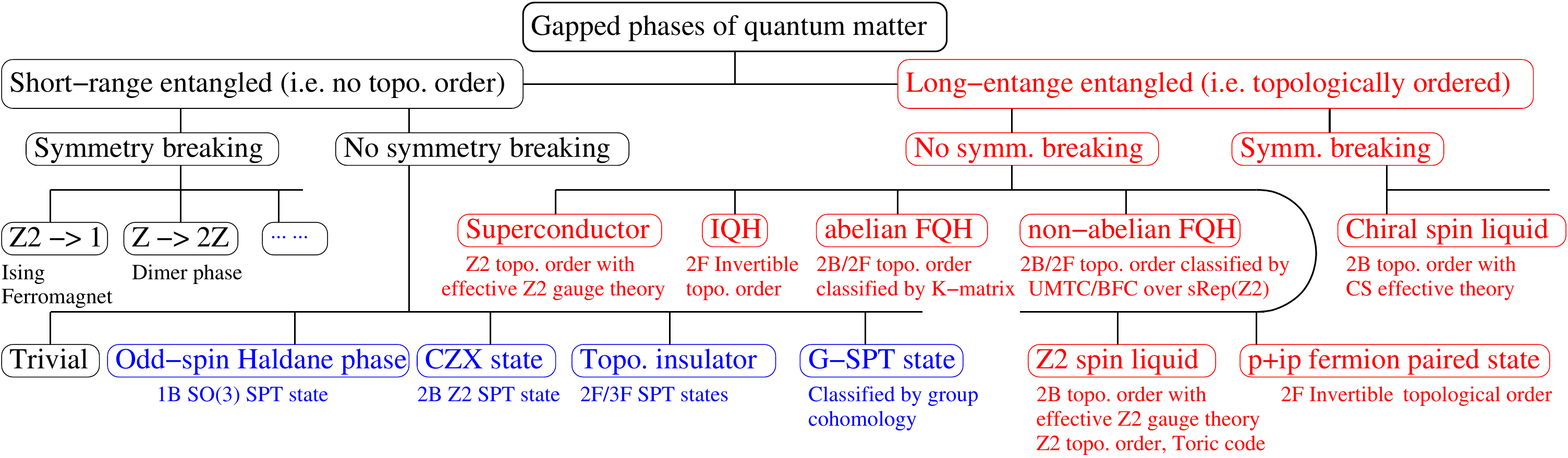}
}
\caption{
Zero temperature phases of matter with energy gap. Only gapped liquids are
listed. The black entries are described by the symmetry breaking.  The colored
entries are phases beyond symmetry breaking: the red ones have topological
order, while the blue are product states with SPT order.  1B refers to 1+1D
bosonic system, 2F 2+1D fermionic system, \etc.  
}
\label{POMT}
\end{figure*}

\subsection{SPT phases: the most trivial non-trivial phases}

In the presence of symmetry, even product states that do not spontaneously
break the symmetry can be non-trivial!  As unentangled states, SPT phases
must be one of the simplest phases of matter. Strictly speaking, SPT phases
have the following defining properties: \emph{(1) different SPT phases can all
be deformed into the same trivial product state smoothly without phase
transition if the deformations break the symmetry, and (2) they cannot be
deformed into each other smoothly without phase transition if the deformations
preserve the symmetry.} 

Such a point of view of SPT phases (\ie stressing their unentangled
trivialness) leads to a fast development of the field. Indeed, only one year
later, a classification \cite{CGW1107,FK1103,SPC1139} of all 1+1D SPT phases
protected by symmetry group $G$ was found in terms of the \emph{projective
representations} of $G$ \cite{PBT1039}.  This leads to a complete
classification of all 1+1D gapped phases of matter.
In particular, the classification predicts that all 1+1D SPT states have
symmetry protected boundary degeneracy which is described by the projective
representations of the symmetry.  Such a measurable character of SPT states
agrees with an earlier result \cite{AKL8799} that spin-1 Haldane phase has a
degenerate spin-1/2 degrees of freedom at a chain end. However,
\textcite{AKL8799} also predicts degenerate spin-1 degrees of freedom at a
boundary of spin-2 Haldane phase, which seems to indicate that the spin-2
Haldane phase is also a non-trivial SPT phase.  Now we know that such spin-1
boundary degrees of freedom are not protected by symmetry since they can be
gapped out by adding several spin-2 degrees of freedom to the boundary without
breaking the $SO(3)$ symmetry. Thus the spin-2 Haldane phase is a trivial SPT
phase \cite{PBT1225}.

Two years later, a systematic theory of bosonic SPT phases in all $d$-dimensions
was developed based on group cohomology $\cH^{d+1}(G,\R/\Z)$ \cite{CGL1172},
and a theory for the boundary of SPT phases was developed by generalizing the
Wess-Zumino-Witten model \cite{WZ7195,W8322}.  One found that SPT phases cannot
have a trivial gapped boundary \cite{CLW1141,VS1306}: the boundary must be
gapless, symmetry breaking, or topologically ordered (if the boundary is
2-dimensional or higher).  Many new phases of matter were discovered
\cite{CLW1141,WS1334,S14054015}.  Shortly after that, a classification of all
bosonic SPT phases in $d$-dimensions was obtained based group cohomology
$\cH^{d+1}(G\times SO_\infty,\R/\Z)$ \cite{W1477}, and based on cobordism
\cite{K1459}.

SPT phases can appear in both bosonic and fermionic systems.  Beside the
bosonic spin-1 Haldane phase that has been realized by experiments
\cite{BMA8671}, the other well known SPT phase is the fermionic topological
insulator protected $G^f= [U^f(1)\rtimes Z_4^T]/Z_2$ symmetry
\cite{KM0502,MB0706,R0922,FKM0703,QHZ0824,CR150503535} that is also realized by
experiments \cite{KWB0766,HH09021356}. Here $Z_4^T$ is the symmetry group
generated by time reversal symmetry, and $U^f(1)$ is the symmetry group for
charge conservation.  The early theories of topological insulators and
topological superconductors \cite{R0664}
are based on non-interacting fermions , in terms of $K$-theory \cite{K0986} or
replica theory \cite{RSF0957}. But many topological insulators and topological
superconductors obtained in non-interacting theory are actually not topological
insulators and topological superconductors in the presence of interactions
\cite{FK1009}.
With interaction,  topological insulators and topological superconductors are
not described by  $K$-theory or replica theory. In this case, they should be
viewed as fermionic SPT phases protected by a symmetry $G^f$, which are
described by group-supercohomology theory of $G^f$\cite{GW1441,GK150505856},
spin cobordism theory \cite{KTT1429}, and/or modular extensions of sRep$(G^f)$
(only in 2+1D) \cite{LW160205946}.

\section{Four revolutions in physics and the second quantum revolution}

It happens several times in the history of physics that a truly new phenomenon
requires a new mathematics to describe it. In fact, the appearance of new
mathematics is a sign of truly new discovery and a revolution in physics. For
example, Newton's theory of mechanics requires calculus to describe the
phenomena of curved motion of particles that was believed to form all matter.
Maxwell theory of electromagnetism reveals a new form of matter: ``wave-like''
matter (the electromagnetic waves and light).  It requires a mathematical
theory of fiber bundle to describe it.  Einstein's theory of general relativity
reveals yet another form of ``wave-like'' matter -- gravitational wave.  It
requires Riemannian geometry. Quantum theory unified ``particle-like'' matter
and ``wave-like'' matter, which requires linear algebra.  Those are four major
revolutions in physics. Newton's and Maxwell's cases are the only two times
when the physicists are ahead of mathematicians, \ie the related mathematics
had not been invented yet when developing those theories.

Right now, we are facing a similar situation.  For a long time, we thought that
different phases of matter are described by symmetry breaking patterns, which
are classified by group theory.  Now we know that there are many new phases of
quantum matter (\ie phases at zero temperature) that are beyond symmetry
breaking (see Fig. \ref{POMT}).  Those new phases are described by the patterns
of many-body quantum entanglement. 

Many-body entanglement is not only the origin of many new states of quantum
matter (such as topological orders), it is also the origin of emergent gauge
fields, and emergent Fermi or fractional statistics.  Those fundamental
properties of nature are simply derived emergent properties the simple bosonic
qubits that form the system.  Recent work found that our empty space itself
might be a particular (long-range) entangled qubit ocean, whose emergent gauge
fields and fermions are the elementary particles in the standard model
\cite{W0303a,W1301,YX14124784}.  Such an emergent picture represents an
unification of gauge interaction and Fermi statistics (not just an unification
of different gauge interactions). It resents  an unification of matter and
information: \emph{it from qubits} (see Fig. \ref{stringnetW}).

Many-body entanglement (\ie topological order) is a truly new phenomenon. It
requires a new mathematics to describe it. But what mathematics describes
pattern of many-body entanglement (and classify topological orders and SPT
orders)?  We have seen that some modern mathematical theories are required.  To
classify SPT orders, we need group cohomology \cite{CGL1172,W1477} or cobordism
\cite{K1459,KTT1429}.  To classify 2+1D bosonic topological orders, we
need unitary modular tensor category \cite{K062,RSW0777,W150605768}.  To
classify 2+1D fermionic topological orders, we need unitary braided fusion
categories over sRep$(Z_2^f)$ \cite{LW150704673}.  But the mathematical theory
to classify topological orders in 3D and beyond is yet to be developed.

Many-body entanglement features new topological states of quantum matter,
topological quantum computation, unification of matter and information, origin
of matter (and even space) from entangled qubits, and new mathematical
foundation of nature.  Thus many-body entanglement (\ie topological order)
represents a second quantum revolution.  Condensed matter physics is not at its
beginning of end. It is just the end of beginning.  It is an exciting time in
physics.

This research was supported by NSF Grant No.  DMR-1506475.

\bibliography{../../bib/wencross,../../bib/all,../../bib/publst}

\end{document}